\documentclass[review]{elsarticle}

\usepackage{hyperref}
\usepackage{longtable}
\usepackage{amsmath,amsthm,amssymb,epsfig,bm,amsmath}
 \usepackage{graphicx}
\usepackage{float}
\usepackage{csquotes}
\usepackage{amsmath}
\usepackage{mhchem}
\usepackage{xparse}
\usepackage{adjustbox}
\usepackage{MnSymbol}
\usepackage{mathdots}
\usepackage{makecell}
\usepackage{listings}
\usepackage{footnote}
\usepackage[flushleft]{threeparttable}
\usepackage{multirow}
\usepackage{relsize}
\usepackage{lettrine}
\usepackage{courier}
\usepackage{color} 
\definecolor{mygreen}{RGB}{28,172,0} 
\definecolor{mylilas}{RGB}{170,55,241}
\usepackage{amsmath,amsthm,amssymb,amsfonts} 
\usepackage{graphicx} 
\usepackage[margin=1in]{geometry} 
\usepackage[yyyymmdd]{datetime}
\usepackage{algorithm}
\usepackage[noend]{algpseudocode}
\usepackage{lipsum}
\usepackage{setspace}
\usepackage{siunitx}
\usepackage{pdfpages}
\usepackage{tikz}
\usetikzlibrary{positioning}
\usepackage{booktabs,caption}
\usepackage[]{hyperref}
\usepackage{soul}
\usepackage{xcolor}

\hypersetup{
 colorlinks  = true, 
 urlcolor   = blue, 
 linkcolor  = black, 
 citecolor  = blue 
}
\usepackage[numbers]{natbib}
\usepackage{color}
\definecolor{codegreen}{rgb}{0,0.6,0}
\definecolor{codegray}{rgb}{0.5,0.5,0.5}
\definecolor{codepurple}{rgb}{0.58,0,0.82}
\definecolor{backcolour}{rgb}{0.95,0.95,0.92}
\lstdefinestyle{mystyle}{
  backgroundcolor=\color{backcolour},  
  commentstyle=\color{codegreen},
  keywordstyle=\color{magenta},
  numberstyle=\tiny\color{codegray},
  stringstyle=\color{codepurple},
  basicstyle=\footnotesize,
  breakatwhitespace=false,     
  breaklines=true,         
  captionpos=b,          
  keepspaces=true,         
  numbers=left,          
  numbersep=5pt,         
  showspaces=false,        
  showstringspaces=false,
  showtabs=false,         
  tabsize=2,
  escapeinside={<@}{@>},
}
 
\usepackage{booktabs} 
\usepackage{siunitx} 
\usepackage{comment}
\usepackage{mathtools} 
\usepackage{gensymb} 
\usepackage{mhchem} 
\usepackage{fixltx2e} 
\usepackage{bbm}

\usepackage{multirow}

\usepackage{fancyhdr} 
\theoremstyle{definition}

\theoremstyle{definition}

\theoremstyle{remark}

\usepackage{soul} 
\soulregister\cite7
\soulregister\ref7
\soulregister\pageref7
\usepackage{bm}

\usepackage{framed} 

\usepackage{multicol} 

\usepackage{nomencl} 
\usepackage{tikz}
\makenomenclature
\usepackage[resetlabels,labeled]{multibib}

\renewcommand*\nompreamble{\begin{multicols}{2}}

\renewcommand*\nompostamble{\end{multicols}}

\usepackage{amssymb}
\usepackage{pifont}
\definecolor{light-gray}{gray}{0.95}

\makeatletter
\AtBeginDocument{\let\hl\@firstofone}
\makeatother

\usepackage[left, pagewise]{lineno}

\journal{a Journal}

\begin{document}


\begin{frontmatter}

\title{\large Multistep Criticality Search and Power Shaping in Microreactors with Reinforcement Learning}

\author{Majdi I. Radaideh$^{a,*}$, Leo Tunkle$^{a}$, Dean Price$^{a}$, Kamal Abdulraheem$^{a}$, Linyu Lin$^{b}$, Moutaz Elias$^{c}$}

\cortext[mycorrespondingauthor]{Corresponding Author: Majdi I. Radaideh (radaideh@umich.edu)}

\address{$^{a}$Department of Nuclear Engineering and Radiological Science, University of Michigan, Ann Arbor, Michigan 48109}
\address{$^{b}$Nuclear Science \& Technology Division, Idaho National Laboratory, Idaho Falls, Idaho 83415}
\address{$^{c}$Lam Research Corporation, Tualatin, Oregon 97062}

\begin{abstract}
\small

Reducing operation and maintenance costs is a key objective for advanced reactors in general and microreactors in particular. To achieve this reduction, developing robust autonomous control algorithms is essential to ensure safe and autonomous reactor operation. Recently, artificial intelligence and machine learning algorithms, specifically reinforcement learning (RL) algorithms, have seen rapid increased application to control problems, such as plasma control in fusion tokamaks and building energy management. In this work, we introduce the use of RL for intelligent control in nuclear microreactors. The RL agent is trained using proximal policy optimization (PPO) and advantage actor-critic (A2C), cutting-edge deep RL techniques, based on a high-fidelity simulation of a microreactor design inspired by the Westinghouse eVinci\textsuperscript{TM} design. We utilized a Serpent model to generate data on drum positions, core criticality, and core power distribution for training a feedforward neural network surrogate model. This surrogate model was then used to guide a PPO and A2C control policies in determining the optimal drum position across various reactor burnup states, ensuring critical core conditions and symmetrical power distribution across all six core portions. The results demonstrate the excellent performance of PPO in identifying optimal drum positions, achieving a hextant power tilt ratio of approximately 1.002 (within the limit of $<$ 1.02) and maintaining criticality within a 10 pcm range. A2C did not provide as competitive of a performance as PPO in terms of performance metrics for all burnup steps considered in the cycle. Additionally, the results highlight the capability of well-trained RL control policies to quickly identify control actions, suggesting a promising approach for enabling real-time autonomous control through digital twins.
\end{abstract}


\end{frontmatter}


\setstretch{1.3}

\section*{Highlights}
\begin{itemize}
    \item Reinforcement learning to perform a criticality search and power control in microreactors was demonstrated.
    \item Proximal policy optimization (PPO) and advantage actor-critic (A2C) are employed.  
    \item Surrogate models for Serpent high-fidelity simulations are used to train reinforcement learning policies. 
    \item Excellent performance was observed for PPO while A2C performance was modest. 
\end{itemize}

\section{Introduction}
\label{sec:intro}

Contemporary nuclear reactor designers are working on microreactors with versatile operating scenarios to operate in remote areas, where automated control policies will be essential to reduce operation and maintenance costs \cite{shirvan2023uo2}. At the same time, advances over the past decade in reinforcement learning (RL) for the optimal control of multiobjective systems have resulted in algorithms that are more efficient and stable than ever \cite{papoudakis2019survey}. However, these algorithms may iterate through millions of  time steps to find reasonable policies \cite{mnih2015ogRL}, and in the case of future advanced reactors, the only available data are through simulations. Simulating each time step accurately enough to train precise and safe control policies must therefore be balanced with minimizing the computational cost of each of these simulations. This paper investigates implementing an accurate and efficient surrogate model to enable RL to produce control policies for microreactors.

Historically developed and tested by finding winning strategies in complex game environments, RL algorithms have seen an exponential increase in their applications to control problems, including the control of error-prone operation phases in pressurized-water reactors \cite{bae2023multiPWR}, plasma control in fusion tokamaks \cite{degrave2022magnetic}, and building energy management \cite{arroyo2022reinforced}, among others. For nuclear reactor design optimization, which can be perceived as a static problem, several studies have demonstrated real-world applications of the deep Q network (DQN) and proximal policy optimization (PPO) RL algorithms in assembly design optimization on small \cite{radaideh2021physics} and large scales \cite{radaideh2021large} along with an open-source implementation framework \cite{radaideh2023neorl}.
In nuclear reactor control, the recent study by Chen et al. \cite{chen2022deep} developed a nonlinear reactivity controller system for a boiling-water reactor using an RL algorithm called a deep deterministic policy gradient. The study found RL more robust compared to another control approach ($H_\infty$) under parameter perturbations and exogenous disturbances. Two other studies focused on using RL to control pressurized-water reactors and obtained their training data from Korea Atomic Energy Research Institute’s compact nuclear simulator, which represents the APR-1400 reactor within certain operating parameters \cite{Kwon1997CNS}. In the first study, Park et al. trained the asynchronous advantage actor-critic (A2C) RL algorithm to adjust two control valves during the heat-up mode to achieve hard-coded pressure and temperature profiles over time \cite{park2022heatup}. In the second, Bae et al. used a combination of the soft actor-critic RL algorithm and the hindsight experience replay technique to learn a control policy for four valves and a heater to achieve pressure, volume, and temperature targets for the heat-up mode, while staying within safe operating limits \cite{bae2023multiPWR}. Both of these studies were only feasible with the availability of the computationally efficient compact nuclear simulator model.

Certain nonnuclear power applications illustrate the potential of RL in our own field. Degrave et al. achieved tokamak feedback control to maintain various, nontrivial plasma shapes using maximum a posteriori policy optimization, another actor-critic RL algorithm \cite{degrave2022magnetic}. Another study by Arroyo et al. combined DQN with model predictive control to learn a strategy for efficient energy use in climate control, subject to considerations of building usage, time-dependent electricity costs, and the thermal inertia of the system \cite{arroyo2022reinforced}. This hybrid controller took advantage of RL to find optimal heating strategies in a complex solution space and model predictive control to maintain constraints without requiring an unreasonable number of training episodes.

This paper addresses a steady-state optimization and control problem for a Westinghouse eVinci\textsuperscript{TM} motivated heat-pipe microreactor design across three different burnup levels. We compare two RL algorithms, A2C and PPO, in learning the optimal progression of control drum positions across different burnup states to achieve criticality while maintaining a balanced power distribution. To enable this comparison on a reasonable timescale, we develop a surrogate model of this system based on high-fidelity Monte Carlo simulations using Serpent \cite{leppanen2015serpent}.

This study offers two key contributions: demonstrating a multiobjective RL achieving a multi-burnup-step criticality search and power flattening in conjunction with a surrogate model and demonstrating that this can be accurately done on short enough timescales to be usable within a digital twin control system.

\begin{table*}[!t]  

\small

\begin{framed}

 \nomenclature{A2C}{Advantage Actor-Critic}
\nomenclature{EMD}{eVinci\textsuperscript{TM} Motivated Design}
\nomenclature{DNN}{Deep Neural Networks}
\nomenclature{DQN}{Deep Q Network}
\nomenclature{HPTR}{Hexant Power Tilt Ratio}
\nomenclature{MDP}{Markov Decision Process}
\nomenclature{MLP}{Multi-Layer Perceptron}
\nomenclature{PPO}{Proximal Policy Optimization}
\nomenclature{RL}{Reinforcement Learning}
\printnomenclature

\end{framed}

\end{table*}

For the remaining sections of this work, Section \ref{sec:background} presents the theoretical background behind the RL algorithms employed in this work. The methodology section, Section \ref{sec:methodology}, details how the surrogate model was trained and utilized as an RL environment as well as how hyperparameters and the reward function were selected for RL algorithms. Section \ref{sec:results} describes the performance of the surrogate model and the RL algorithms used. Further context, limitations of this work, and avenues for further investigation are found in Sections \ref{sec:discussion} and \ref{sec:conclusion}, respectively.

\section{Reinforcement Learning Theory}
\label{sec:background}
RL aims to find a strategy that chooses the best action in a given situation that will maximize a long-term reward. Within the RL field, this strategy is called a policy and the situation is a state. An agent performs actions according to the policy within an environment. Formalism in this field begins by idealizing RL as a Markov decision process (MDP), for which an action $A_t$ performed in state $S_t$ results in a new state $S_{t+1}$ with a certain probability along with a corresponding reward $R_{t+1}$. A trajectory of states and actions looks like: $S_0, A_0, R_1, S_1, A_1, R_2, S_2, A_2, R_3, S_3$, etc.

A finite trajectory is called an episode. For nonfinite tasks, the trajectory must be truncated at some point to create an episode. Key to an MDP is the Markov property, which requires that each state has enough information for the rest of the trajectory to proceed without referencing previous states. For instance, the trajectory progression after $A_2$ is taken in $S_2$ should not need to account for $S_0$ or $S_1$. This property motivates and allows for several recursive relationships to be defined. The first of these is the return, commonly denoted $G$ and representing the total rewards that will accumulate from the trajectory at a given state.

\begin{align}
    G_t &= R_{t+1} + \gamma R_{t+2}  + \gamma^2R_{t+3} + \gamma^3R_{t+4} + \ldots \notag\\
    &= R_{t+1} + \gamma (R_{t+2}  + \gamma R_{t+3} + \gamma^2R_{t+4} + \ldots) \\
    &= R_{t+1} + \gamma G_{t+1} \notag
\end{align}

\noindent where $\gamma$ is the discount factor between 0 and 1, but typically greater than .9, and is introduced to determine whether the policy should focus on immediate or long-term rewards. The return expected from a state $s$ following policy $\pi$ is called the value function:

\begin{equation}
    v_\pi(s) = \mathbb{E}_\pi[G_t | S_t = s]
\end{equation}

By convention, uppercase is used to denote random variables while lowercase represents an observation of the variable. A similar action-value function describes the expected return if action $a$ (without regard to $\pi$) is taken at state $s$ after which policy $\pi$ is followed:

\begin{equation}
    q_\pi(s,a) = \mathbb{E}_\pi[G_t | S_t = s, A_t = a]
\end{equation}

While it is possible to have a deterministic policy, $\pi(a|s)$ is interpreted as the probability of taking action $a$ in state $s$. For any problem, there exists a policy that is better than or equal to every other policy. This is known as an optimal policy. The details of this derivation are outside the scope of this paper, but for an optimal policy, the so-called Bellman optimality equations for the optimal value and action-value are:

\begin{align}
    v_*(s) &= \mathop{max}_a \mathop{\sum}_{s',r} p(s',r | s,a)[r + \gamma v_*(s')] \notag\\
    q_*(s,a) &= \mathop{\sum}_{s',r} p(s',r | s,a)[r + \gamma \mathop{max}_{a'} q_*(s',a')]
\end{align}

Considering every state, with terminal states of finite trajectories defined as having a value of zero, this becomes a system of equations with one equation and one unknown per state or state-action pair. Thus $v_*$ and $q_*$ can theoretically be directly solved for. Once these are known, the optimal policy is simply a matter of choosing the action that maximizes $q_*$ in a given state. However, for all but the simplest problems, this is intractable. Most techniques in RL, therefore, make various approximations to these optimality equations to hopefully find nonoptimal but good policies. The two major categories of these techniques are value- and policy-based methods.

Value-based methods attempt to approximate the value of each state and are the most obvious extension of the Bellman optimality equations. With unlimited time, Monte Carlo methods can calculate returns in trajectory after trajectory and get successively better estimates of each value. However, complex problems tend to have state spaces that are far too large to hold in computer memory, so using a functional approximation of the state value is more practical. An artificial neural network, which acts as a universal function approximator that can improve with experience through backpropagation, serves this purpose well. One of the first major successes of this approach was TD-Gammon, which held its own against backgammon grandmasters \cite{tesauro1995backgammon}. Deep Q learning uses a convolutional neural network and was able to outperform human players at certain Atari games when first published \cite{mnih2013dqn}.

Policy-based methods directly select actions without reference to the value of a state. They are parameterized such that policy parameters, denoted $\theta$, can be updated with the gradient of some performance measure to improve the policy through experience. The canonical policy gradient method, REINFORCE \cite{williams1992reinforce}, updates $\theta$ in the direction of $\nabla_\theta \log \pi(a_t | s_t;\theta) G_t$.

A pure policy gradient method tends to be very slow to converge to a good policy due to high variance in how large these updates are. By subtracting an estimate of the state value, V(s), from the return, $G_t$, this variance can be greatly reduced. Now $\theta$ gets updated in the direction of $\nabla_\theta \log \pi(a_t | s_t;\theta)(G_t - V(s_t))$. This combination of policy- and value-based techniques is known as an actor-critic method. The critic improves its ability to value states, and this value is used within the gradient descent learning process to improve the actor’s policy. $G_t – V(s_t)$ is an estimator of the advantage, defined as $\mathcal{A}(a_t, s_t) = Q(a_t, s_t) – V(s_t)$.

 Adding an estimate of the advantage to the policy gradient serves to increase the probability of above-average actions and decrease the probability of below-average actions. Adding an entropy term to the policy gradient to prevent early convergence results in the A2C RL algorithm. Mnih et al. showed that running multiple actor-learners in parallel and merging their updates at regular intervals reduced the training time, stabilized the learning process, and improved the resulting policies \cite{mnih2016a3c}. In this paper, we take advantage of this parallelism and use the Stable Baselines3 \cite{sb3} implementation of A2C.

Like all machine learning methods, A2C requires careful hyperparameter tuning. Additionally, while having multiple actor-learners can reduce the frequency of overly large gradient updates through cancellation, the problem is not eliminated completely. In an early attempt to address this issue, Schulman et al. introduced the concept of a trust policy region, in which the policy update is constrained in size by ensuring the average Kullback-Leibler divergence per state between the new and old policies is bounded \cite{schulman2015trpo}. While this improved stability and reduced the need for hyperparameter tuning, the implementation was complex and incompatible with commonly used RL architectures.

Later, Schulman et al. published a simpler, more general algorithm with the same benefits called PPO \cite{schulman2017ppo}. Instead of using a complex trust region, this uses a simple clipping function with parameter $\epsilon$, which prevents the probability ratio of a new policy to the old policy during gradient updates from being below $1-\epsilon$ or above $1+\epsilon$. Additionally, where A2C estimates the advantage with $G_t – V(s_t)$, PPO uses a generalized advantage estimation, which is tunable to trade-off between bias and variance in the estimate \cite{schulman2018gae}. Several standard RL benchmark problems demonstrated that PPO usually outperformed other algorithms, even when they had hyperparameter tuning for the specific problem and PPO did not. Since PPO is generally implemented on top of A2C, including in the Stable Baselines3 implementation we use, we again take full advantage of parallel actor-learners.
 
\section{Methodology}
\label{sec:methodology}




\subsection{Heat-Pipe Microreactor Model}
\begin{figure}[htb!]
    \centering
    \includegraphics[width=\textwidth]{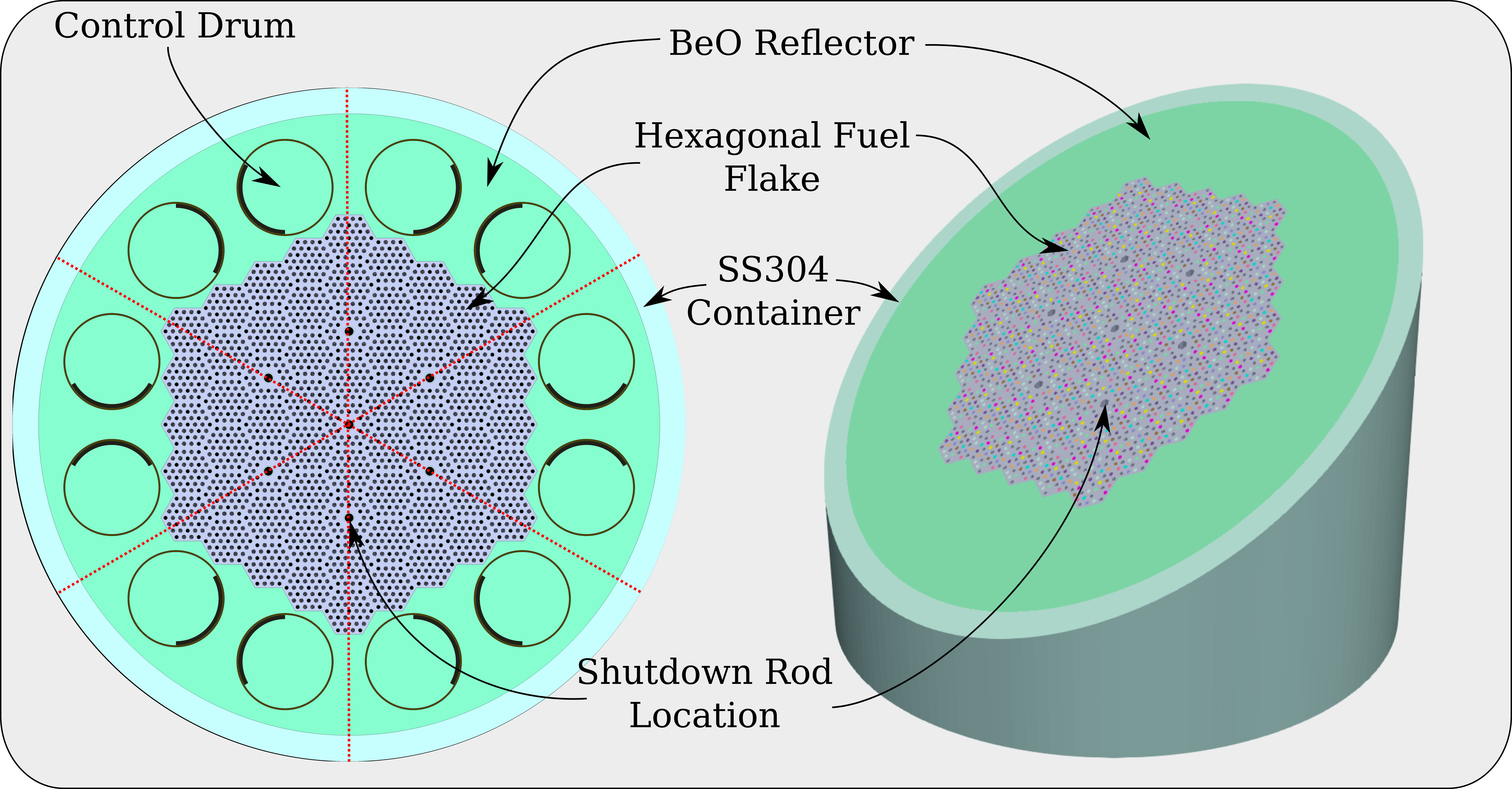}
    \caption{EMD with control drums oriented at 90$^\circ$, shown with the slice at the core midplane (left) and a 3D rendering with a diagonal cut (right). Hexant divisions are shown with dotted red lines. Control drums are not rendered in the 3D depiction.}
    \label{fig:EMD}
\end{figure}

In this study, Serpent \cite{leppanen2015serpent} is used to calculate neutronic quantities associated with a heat-pipe microreactor design motivated by the Westinghouse eVinci\textsuperscript{TM} design.
The particular design used in this study will be referred to as the eVinci\textsuperscript{TM} motivated design (EMD) and is selected because its material and geometrical specifications are publicly available \cite{price2023thermal}.
The EMD is designed to operate at 4 MWth with a 4 year lifetime, and it has 12 control drums, which are used to control reactivity.
An illustration of the core is provided in Figure \ref{fig:EMD} where the Serpent-rendered core geometry at the midplane is provided alongside a three dimensional (3D) rendering of the geometry.
In this figure, the 12 control drums are shown at a 90$^\circ$ orientation in the 2D Serpent-rendered core geometry.
These control drums can be rotated to manipulate the reactivity and power distribution of the core.

\subsection{Surrogate Modeling}
\label{sec:surr}
In this analysis, calculating neutronic core characteristics using the Serpent neutronics code would be too computationally expensive for the large number of evaluations required for the RL algorithm.
Ideally, the control problem would be integrated into a digital twin that would be fed with real-time sensor data---there would be no need for any direct calculation of these quantities.
Therefore, a set of surrogate models were created to predict core properties given a set of control drum positions at a particular cycle time.
The relevant core properties for the control problems considered here are the core multiplicity and hexant power fractions.
The hexant power fractions are a set of six quantities that describe the fraction of total power that comes from six portions of the core.
The divisions used to calculate these hexant powers are shown in Figure \ref{fig:EMD} with red dotted lines.
The control drums within each hexant are considered to move together, in opposite directions, such that the positions of the 12 total control drums can be described with a set of six control drum angles.

A nominal core depletion calculation is run in Serpent with all control drums facing 90$^\circ$.
Then, three fuel composition states can be identified: 0 YR corresponds to the beginning of life fuel composition when the core is loaded with fresh fuel, 2 YR corresponds to the core fuel composition after 2 years of operation, and 4 YR corresponds to the core fuel composition after 4 years of operation.
Serpent is used to calculate core multiplicities and hexant powers for 250 control drum positions for each of these three fuel compositions for a total of 750 calculations.
The Monte Carlo sampling statistics were selected such that the final uncertainty in core multiplicity was estimated to be about 9 pcm.
Within each fuel composition, symmetries in the core geometry can be exploited to turn these 250 calculations into 3,000 data points, which can be used to train and evaluate the surrogate models.
These 3,000 data points are divided into a training data set, used to calculate weights and biases associated with the deep neural network (DNN) surrogate model, a validation set, used to select an optimal network architecture, and a testing set, used to evaluate the final performance of the DNN, with a 70\%/15\%/15\% split, respectively.
The divisions between these sets did not split data points, which may have come from the same Serpent calculation (i.e., symmetries were applied after data are divided among the three data sets).

For each of the 0 YR, 2 YR, and 4 YR, a single DNN with six inputs (control drum position) and seven outputs ($k_{eff}$, $P_1$, $P_2$, ..., $P_6$) was created to simultaneously predict core multiplicity and six fractional hexant powers given a set of six control drum angles.
For architecture optimization, the number of hidden layers, nodes per hidden layer and learning rate are to minimize the mean-squared error when predicting the validation data set independently for each of the three DNNs corresponding to the 0 YR, 2 YR, and 4 YR cases.
A constant number of nodes per hidden layer is used to simplify the optimization process.
Bayesian optimization is used for this mixed-integer optimization problem with an allowance of 30 search trials.
The results from this training process, including performance measures of the surrogate models, are discussed in Section \ref{sec:surr_perf}.

\subsection{RL Environment}

\begin{figure}[htb!]
    \centering
    \includegraphics[width=.8\textwidth]{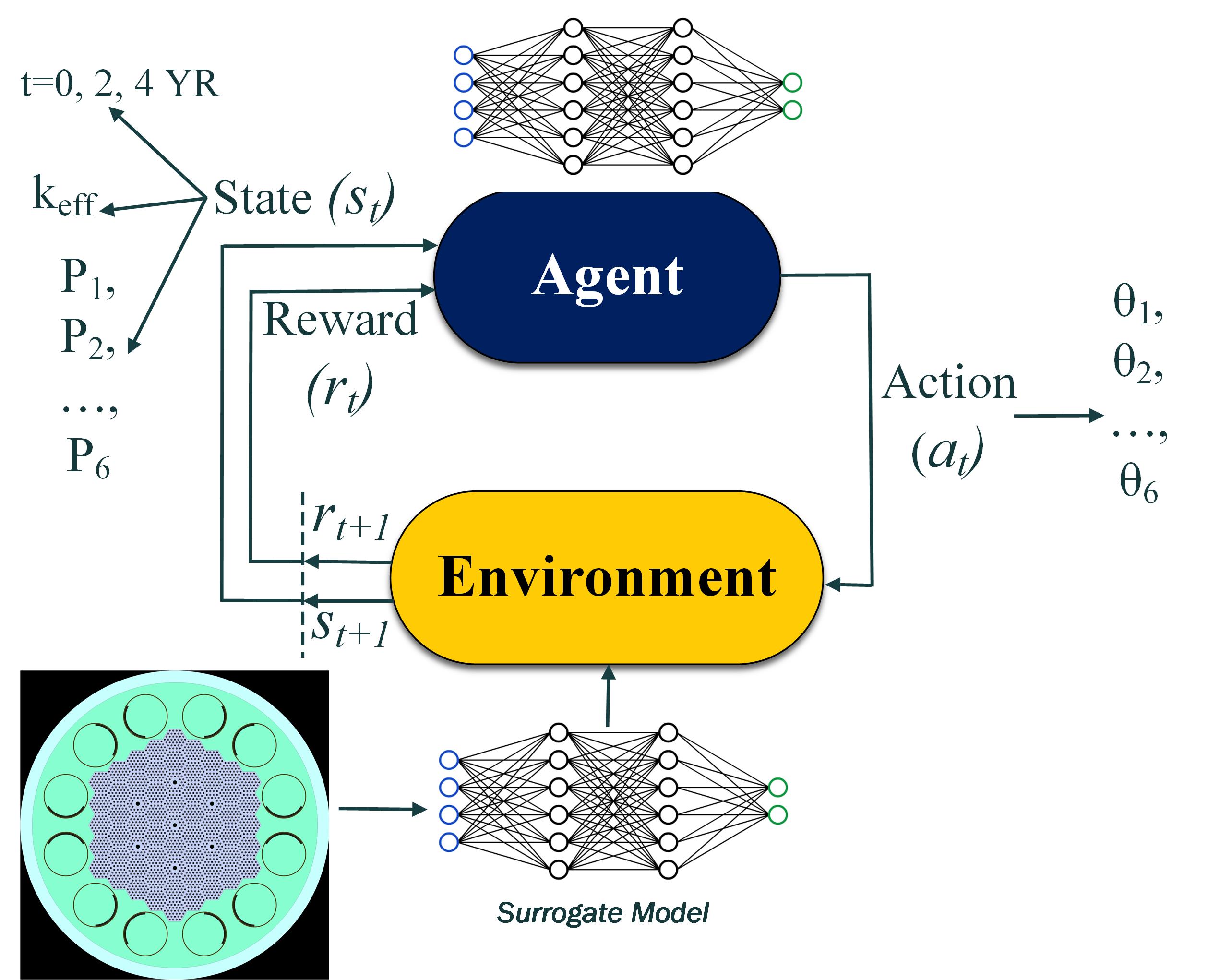}
    \caption{The RL control framework adopted in this work.}
    \label{fig:rlcontrol}
\end{figure}

Figure \ref{fig:rlcontrol} illustrates the RL framework used in this work. Originally developed by OpenAI, ``Gym'' is the most widely used Python library for defining an environment \cite{openaigym}. We use Gym Version 0.19.0. The details of our environment, which consults the surrogate model to compute rewards, are:

\begin{itemize}
    \item Action space ($a_t$): the possible control actions upon the environment. While the control drum orientation in each of the six reactor hextants can physically take any value between 0$\degree$ and 360$\degree$, we reduced the search space to only discrete integer values between 0$\degree$ and 180$\degree$. Here we take advantage of the symmetry of the drum reactivity worth, which is similar from 0$\degree$ to 180$\degree$ and 180$\degree$ to 360$\degree$. The discretization of the action space will facilitate the training process and reduce training time. Following Gym’s nomenclature, we use a multidiscrete object of Size 6 bounded by 0 and 181 (Index 181 is not reached by the agent).
    \item State space ($s_t$): the possible information given to the agent to base its next action upon. There are three parts shaping the state space: the current time step (0 YR, 2 YR, 4 YR), the core multiplicity ($k_{eff}$), the power fraction of each hextant ($P_1, P_2, \ldots, P_6$).
    \item Reward ($r_t$): designed to encourage power symmetry in the six hextants and a critical core at each burnup time step. The reward is calculated using $k_{eff}$ and power fraction outputs from our surrogate model as follows:
    \begin{equation}
    \label{eq:rwd}
        r_t^0 = \frac{1}{f_1^t+f_2^t+ f_3^t}, \quad t=0, 2, 4 \ YR,  
    \end{equation}
    where 
    \begin{equation}
    \label{eq:f1}
        f_1^t = \frac{1}{6}\sum_{i=1}^6 |P_i - 0.166667|,
    \end{equation}
    \begin{equation}
        f_2^t = |k_{eff}^t - 1.00000|.
    \end{equation}
    \begin{equation}
        f_3^t = \sqrt{\frac{\sum_{i=1}^6 (P_i - \bar{P})^2}{6}}
    \end{equation}
    where $\bar{P}$ is the average of the six hextant power fractions. The third term $f_3^t$ was not reported in our PHYSOR paper \cite{radaideh2024demonstration} since its impact on the learning process was negligible. However, we still report that here for completeness. Since RL maximizes rewards, the reciprocal is taken in Eq.\eqref{eq:rwd} to convert a min to max problem without changing the sign. 
\end{itemize}

Episodes are very short and defined as three successive burnup time steps in which the agent attempts to input the optimal angles for each one. To avoid plotting all six power fractions, we define a hextant power tilt ratio (HPTR) as follows:

\begin{equation}
    HPTR = \frac{6 \times max(P_1, P_2, ..., P_6)}{\sum_{i=1}^6 P_i}
\end{equation}

This expression is based on a commonly used equivalent for cores with four quadrants and is only used for plotting and monitoring purposes.

The episode always starts with the reactor at time $t=0$ YR with random $k_{eff}$ and $P_i$ values, which represent the initial state $s_1$. The agent takes the action ($a_1$) and calculates the reward ($r_1$). The state is then updated to $s_2$, which is equivalent to $t=2$ YR. The agent then repeats the same process to get to $s_3$ at $t=4$ YR. The goal is to maximize the reward of the episode by taking the optimal action in the three time steps. After the episode terminates, a new episode starts and the process is repeated many times until the agent learns how to play the game. Due to the sequential nature of the episode, we modify the definition of the reward function in Eq.\eqref{eq:rwd} to:

\begin{align}
\label{eq:rwd2}
\begin{split}
 r_1 &= r_1^0, \\
 r_2 &= MEAN(r_1^0, r_2^0) - STD (r_1^0, r_2^0), \\
 r_3 &= MEAN(r_1^0, r_2^0, r_3^0) - STD (r_1^0, r_2^0, r_3^0) .
\end{split}
\end{align}

We observed that this adjustment to the reward function enables the agent to prioritize learning across all three burnup steps simultaneously, rather than becoming biased and focusing on increasing the reward for a single, potentially easier-to-optimize burnup step, which could occur if the reward function in Eq.\eqref{eq:rwd} was issued as is. To achieve an optimal reward, the mean reward across the three burnup steps must be maximized, while their standard deviation must be minimized.

\subsection{RL Hyperparameters}
\label{sec:hyperparameters}
As with most machine learning methods, RL algorithms require problem-specific hyperparameter tuning. For this work, we performed a grid search among the following hyperparameters described further below: the value coefficient, the entropy coefficient, $N_{steps}$, the maximum norm of the gradient, and the PPO clipping parameter.

While the actor and critic components of A2C and PPO can be approximated by two separate neural networks, most implementations, including that of Stable Baselines3, employ a single neural network with outputs for both the value and policy. To improve both at once, the following objective is maximized:

\begin{equation}
    L^{total}(\theta) = L^{policy}(\theta) - c_1 L^V(\theta) + c_2 H[\pi_\theta](s_t)
\end{equation}

\begin{itemize}
    \item The value coefficient, $c_1$, tells the neural network how much importance to give the value portion of the output when updating parameters through backpropagation. Due to the negative sign in front of the value coefficient, the higher it is, the more it hinders the maximization of the combined objective function. A value of 0.5 would weight the policy and value components of the objective function equally, while a value of 0.75 would weight the value component more.
    \item The entropy coefficient, $c_2$, determines the contribution of the entropy term, which prevents the early optimization of the policy before it has fully explored the environment.
    \item $N_{steps}$ determines how many steps to learn from before updating the policy. $N_{steps}=3$, for instance, would update the policy after every episode, which could be too soon and result in an unstable policy. $N_{steps}=3000$ would update the policy after 1,000 episodes, which could be too long to wait and waste computing resources and experience for little improvement.
    \item The maximum gradient norm imposes a limit on how large the policy gradient can be. This prevents the issue of exploding gradients, which causes the policy update step to overshoot and result in an unstable policy. Since PPO is known to be less sensitive to hyperparameters, we chose to only tune this for A2C in favor of tuning PPO’s clipping parameter, $\epsilon$.
    \item The clipping parameter $\epsilon$ is described in Section \ref{sec:methodology} and is embedded in $L^{policy}(\theta)$. It does not consider the gradient, only the ratio of action probabilities between the current policy and the new policy proposed after a gradient update.
\end{itemize}

\subsection{RL Simulation Details}
All runs were completed with 20 processors running agent-learners in parallel. The surrogate model used Keras-2.6.0\cite{chollet2015keras}, the learning environment was created with Gym-0.19.0, and the A2C and PPO implementations were from Stable Baselines3-1.1.0. The default actor-critic model was used for each implementation and consists of a feedforward neural network with two hidden layers of 64 nodes each. The training process was limited to 1.2 million time steps, which kept the overall training time around 1 hour per run.

\section{Results}
\label{sec:results}

The results of this paper are presented in Subsections \ref{sec:surr_perf}-\ref{sec:rl_test}. First, the surrogate model of the Serpent simulation is trained and validated. The surrogate model is then used as an environment to train the A2C and PPO RL algorithms. Lastly, the trained policy is tested to ensure it can identify the optimal drum positions.

\subsection{Surrogate Model Performance}
\label{sec:surr_perf}
In total, three different surrogate models for 0 YR, 2 YR, and 4 YR are used to supply core multiplicity and hexant powers (as fractions of total power) for a given set of control drum angles.
In order to generate a network architecture that can adequately capture the relationships between control drum angles and core properties, a hyperparameter optimization routine is run where the learning rate, number of layers, and nodes per layer are selected to ensure strong performance.
After the DNNs are trained with these architectures, their performance is evaluated on a set of 456 testing data points.
The performance metrics of these DNNs on this testing set, as well as their network architectures, are shown in Table \ref{tab:surr_perf}.
For the hexant powers, the surrogate model produces six different values, each corresponding to a fractional power in the corresponding section of the core.
For brevity, the performance metrics provided in Table \ref{tab:surr_perf} pertaining to these hexant powers have been averaged such that only a single set of performance metrics is provided.
There is relatively little variation in model performance when predicting the powers in each hexant because symmetry is used to expand the training data.

Besides neural networks, we have trained other model types like linear regression and random forests. Linear regression provided a weak performance while random forests provided $R^2 \sim 0.8$. We continued with neural networks due to their better performance. 

\begin{table}[htb!]
\centering
\caption{Network architecture and performance of surrogate models evaluated on the testing set. Mean absolute error (MAE) and coefficient of determination (R$^2$) are shown when using the surrogate model to predict core multiplicity ($k$) and fractional hexant powers ($P$). }
\label{tab:surr_perf}
\begin{tabular}{|l|l|l|l|}
\hline
\textbf{Parameter}       & \textbf{0 YR}                   & \textbf{2 YR}                   & \textbf{4 YR}                   \\ \hline
\textbf{Learning Rate}   & 3 $\times~ 10^{-3}$    & 2 $\times~10^{-3}$    & 2 $\times~10^{-4}$    \\ \hline
\textbf{Number of Layers}     & 5                     & 5                     & 7                     \\ \hline
\textbf{Nodes per Layer} & 150                   & 200                   & 191                   \\ \hline
\textbf{$k$: MAE}        & 85 pcm                & 83 pcm                & 79 pcm                \\ \hline
\textbf{$k$: R$^2$}      & 0.996                 & 0.996                 & 0.998                 \\ \hline
\textbf{$P$: MAE}        & 2.32 $\times~10^{-4}$ & 2.33 $\times~10^{-4}$ & 2.29 $\times~10^{-4}$ \\ \hline
\textbf{$P$: R$^2$}      & 0.999                 & 0.999                 & 0.999                 \\ \hline
\end{tabular}
\end{table}

\subsection{RL Training Results}
\label{sec:rl_train}
After a grid search for each algorithm, the hyperparameters described in Section \ref{sec:hyperparameters} were selected for each algorithm. A2C performed best with a value coefficient of 0.5 ($c_1=0.5$), an entropy coefficient of 0.02 ($c_2=0.02$), 200 steps before updating the policy ($N_{steps} = 200$), and a maximum gradient norm of 5. For PPO, these were $c_1=0.75$, $c_2=0.01$, $N_{steps} = 300$, along with a clipping parameter of 0.4 ($\epsilon=0.4$).

While a small amount of entropy gave better results compared to the no entropy default, higher entropy coefficients (such as $c_1$ = 0.1 or 0.2) yielded worse performance in each case, with training failing to converge to a good policy within the 1.2 million time steps. Since both algorithms were run with 20 processors, the effective number of steps before a policy update was 20 $\times$ 200 = 4,000 time steps and 20 $\times$ 300 = 6,000 time steps for A2C and PPO, respectively. A2C was quite sensitive to changes in $N_{steps}$, with values of 300 or higher resulting in rapidly worse performance.

\begin{figure}[htb!]
    \centering
    \includegraphics[width=\textwidth]{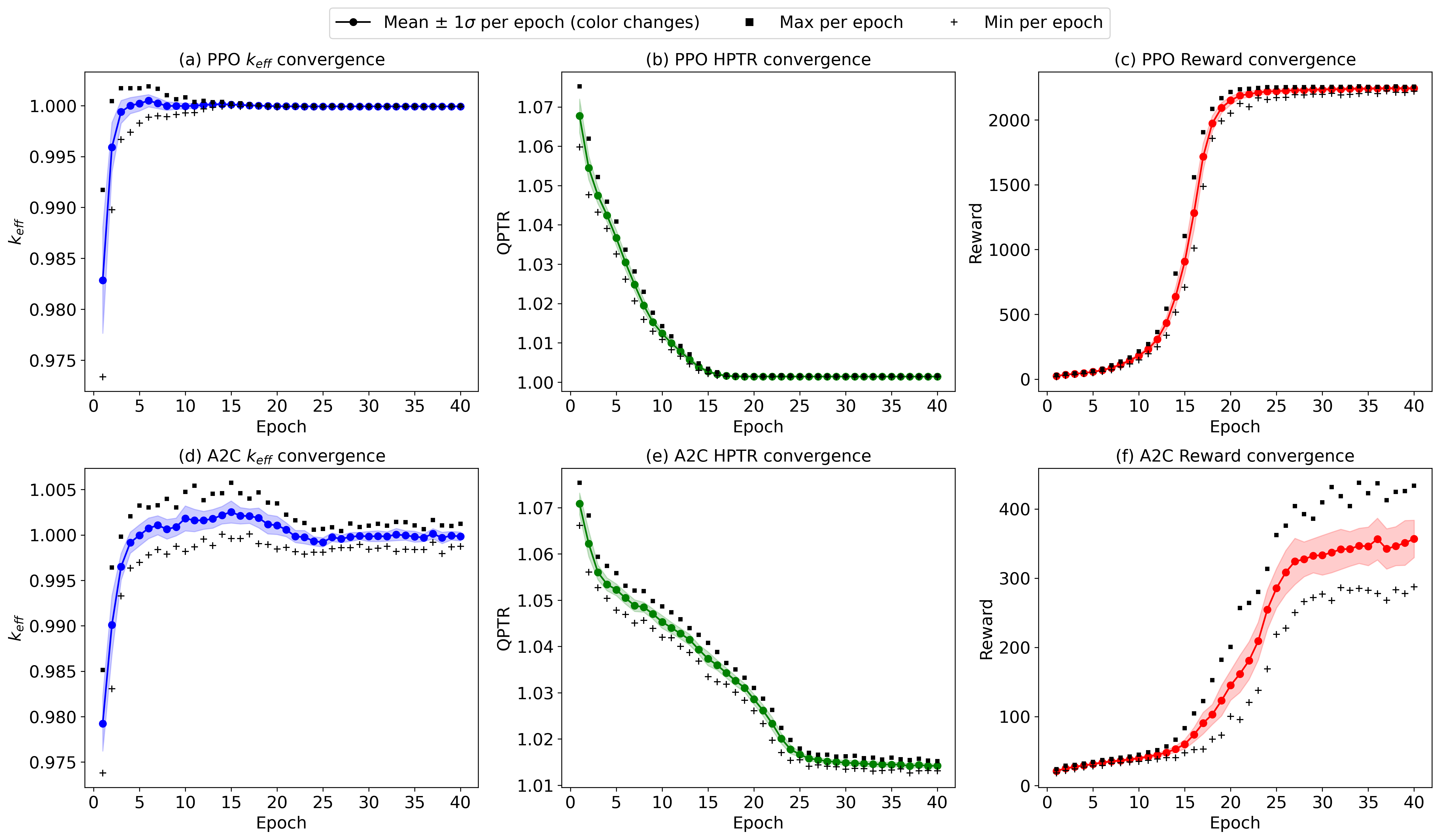}
    \caption{Progression of PPO and A2C training as a function of the number of epochs. Each epoch comprises 30,000 time steps for which the statistics (mean, std, max, min) are derived. The top row shows the convergence of $k_{eff}$, HPTR, and the reward value as indicated in Eq.\eqref{eq:rwd} for PPO while the bottom row shows these plots for A2C. All values shown are averaged over the three reactor states and 20 parallel agents. Also, note that the reward scale in this plot is slightly different from Figure 3 of \cite{radaideh2024demonstration} (see the main text for explanation).}
    \label{fig:rl_plt}
\end{figure}

Training progress plots for both algorithms are shown in Figure \ref{fig:rl_plt}. The subplots show how $k_{eff}$, HPTR, and the reward converge to optimal values. Each of these values were averaged over each episode and over the parallel processors, so they represent the average performance of the 20 parallel agents over the three time steps. Additionally, each point plotted represents statistics from 30,000 aggregated time steps (10,000 episodes), which we define as an epoch.

It is important to note that the reward scale for PPO in Figure \ref{fig:rl_plt}(c) differs slightly from the one originally published in our PHYSOR conference paper \cite{radaideh2024demonstration}. The authors discovered a typo in the source code regarding the use of the absolute value after the mean in Eq.\eqref{eq:f1}, which should be reversed. This did not affect the training process or the performance of PPO in any way. However, the reward scale in Figure 3 of \cite{radaideh2024demonstration} differs from Figure \ref{fig:rl_plt}(c), even though both reward scales yield similar HPTR and $k_{eff}$ values. This note is provided to avoid confusing the reader about having similar reward formulations between the two studies that yielded two different reward values. 

For both algorithms, $k_{eff}$ is rapidly improved over the first five epochs, while power balancing takes longer to improve. PPO exhibits policy convergence within 25 epochs. Meanwhile, the high variability in rewards obtained by successive policies in later epochs, together with the lower magnitude of rewards compared to PPO, indicate that A2C is unable to converge to an optimal policy within the time step training window.

\subsection{RL Testing Results}
\label{sec:rl_test}
Following RL training, we proceeded to evaluate performance with the most recent versions of each policy, asking both the A2C and PPO policies to provide optimal drum positions for each burnup time step. These are displayed in Table \ref{tab:rl_test}. For every reactor state, the PPO policy was able to find accurate drum angles. This is evident from the excellent $k_{eff}$ and HPTR values achieved compared to the target values. 

A2C on the other hand is still providing reliable results for time 0 YR and 4 YR as shown by the $k_{eff}$ and HPTR values for these burnup steps and how close the drum angles identified by A2C compared to PPO. However, A2C has struggled to figure out drum orientation for $t=$2 YR as $k_{eff}$ deviates about 500 pcm from criticality and HPTR is slightly above the constraint of 1.02. This implies that the lower reward values illustrated in Figure \ref{fig:rl_plt}(d) for A2C can be attributed to $t=$2 YR performance metrics, which may not be obvious due to averaging the three bunrup steps. Nevertheless, unlike PPO, the larger standard deviation observed for A2C rewards and the larger gap between the max and min rewards indicate that not all burnup steps have converged to satisfactory results.  


\begin{table}[htbp]
  \centering
  \caption{Testing results of the trained PPO and A2C policies for each burnup state.}
\begin{tabular}{|l|l|l|l|l|l|l|l|l|l|}
\hline
\textbf{Algorithm} & \textbf{Time (YR)} & \pmb{$\theta_1$} & \pmb{$\theta_2$} & \pmb{$\theta_3$} & \pmb{$\theta_4$} & \pmb{$\theta_5$} & \pmb{$\theta_6$} & \pmb{$k_{eff}$} & \textbf{HPTR}     \\ \hline
PPO      & 0         & 91         & 89         & 95         & 88         & 93         & 89         & 1.00002   & 1.002679 \\ \hline
PPO      & 2         & 114        & 113        & 112        & 111        & 114        & 113        & 0.99993   & 1.001708 \\ \hline
PPO      & 4         & 138        & 136        & 129        & 141        & 134        & 138        & 1.00000   & 1.002824 \\ \hline
A2C      & 0         & 92         & 95         & 85         & 90         & 88         & 94         & 1.00008   & 1.005743 \\ \hline
A2C      & 2         & 92        & 132        & 85        & 139        & 82        & 135        & 0.99502   & 1.027286 \\ \hline
A2C      & 4         & 140        & 132        & 133        & 139        & 135        & 135        & 0.99998   & 1.003103 \\ \hline
\end{tabular}
  \label{tab:rl_test}%
\end{table}%

%

\section{Discussions and Implications}
\label{sec:discussion}
Each surrogate model was trained on 2,100 data points. For a six dimensional input space and seven dimensional output space, this is not much data, especially when considering the complexity of the DNNs used. The total number of learnable parameters for each DNN surrogate was around 70,000, 120,000, and 190,000 for the 0, 2, and 4 YR burnup models, respectively. These high numbers of parameters resulted through our architecture optimization process and were necessary to keep errors $k_{eff}$ errors low enough to be useful. A high number of parameters comes with the danger of overfitting to the training data; however, the test results show excellent performance, demonstrating that this did not occur. This may indicate that the underlying relationship between drum angles and $k_{eff}$ and hextant power fractions is relatively simple.

A limitation of our method of surrogate modeling is that it requires a separate model for each burnup time. This limited our RL work to very short episode lengths of three burnup time steps. However, without this highly accurate surrogate model, running 1.2 million separate Serpent calculations would be necessary, which would make sample-hungry RL impractical. Scaling the DNN model to capture more burnup steps, and even using one model for all burnup steps, is feasible if more data can be generated from Serpent. In addition, capturing model uncertainty through models like Gaussian processes \cite{radaideh2020surrogate} might become necessary when model accuracy becomes limited. 

Both A2C and PPO learned how to take actions leading to criticality and balanced hextant powers. In each case, optimizing for criticality was easier than learning to act symmetrically. That there are far more critical states than symmetric power states explains this, since each drum may move independently to achieve the criticality, but must move together for symmetry to hold. In fact, it seems that, with both algorithms, it took around six times longer for the HPTR to reach its lowest levels than it did for $k_{eff}$ to get close to 1, which corresponds with the relative number of satisfactory states for each objective.

In terms of overall performance, however, PPO was far more robust than A2C within the training time frame. A2C remained unstable even as it improved its policy. In sharp contrast with PPO, which became more stable with time, A2C maintained a level of instability proportional to the mean reward and failed to converge within the training time for $t = 2 YR$. Also, HPTR values achieved by PPO are still lower than A2C for all three burnup steps. These results are consistent with the Stable Baselines3 implementation of PPO as a smoother variant of A2C due to the addition of generalized advantage estimation and a clipping term to prevent overly large policy updates.

Ultimately, these algorithms learned something that any nuclear engineer already knows: a balanced power profile can be achieved through symmetric drum angles and there should be a single, symmetric criticality position for a given material and temperature makeup. However, they did this with only a simple reward function to guide it through a thirteen dimensional space (six angle actions, six power fractions, and $k_{eff}$) within 1 hour. For people in the RL field, this result is probably unsurprising, but RL applications for nuclear reactor control are sparse, with the three examples described in Section \ref{sec:background} by Chen \cite{chen2022deep}, Park \cite{park2022heatup}, and Bae \cite{bae2023multiPWR} being the only ones we could find. None of these looked at microreactors.

While RL training requires about 60 minutes to complete, it takes only around $\sim0.025 \ s$ for a trained policy to provide an optimal action for a given burnup time. This rapid decision capability could one day help enable real-time autonomous control. However, the black-box nature of the underlying models would mandate intense scrutiny for this purpose. Even so, while neural network policies are not interpretable, the actions they take can be studied. Especially with new microreactor designs that have not yet been built and so have no operational experience, RL has the potential to anticipate issues and opportunities in reactor control.

The natural progression of this work would involve applying this methodology for real-time control in transient calculations while adhering to the physical limitations of drum movements (such as maximum drum speed and reactivity insertion per degree of movement). For instance, performing transient simulations in load-following scenarios using point kinetics models will allow RL to receive feedback from each degree of rotation based on the reactivity inserted and the resulting power changes. Since load-following typically occurs over relatively long periods compared to reactor response times (e.g., 10--20 minutes), RL algorithms can adapt to system feedback while adjusting the drums to achieve the desired power level. Future efforts could also include comparisons with traditional time-series models like long short-term memory, which have been demonstrated for transient modeling in nuclear systems \cite{radaideh2020neural}.
\section{Conclusions}
\label{sec:conclusion}

In this work, we have demonstrated an initial application of intelligent control using RL for nuclear heat pipe microreactors. The RL agent was trained using PPO and A2C, advanced deep RL techniques, based on a high-fidelity simulation of a microreactor design inspired by the Westinghouse eVinci\textsuperscript{TM} design. A Serpent model provided high-fidelity data on drum positions, core criticality, and core power distribution for training a surrogate model. This surrogate model was then used to train a PPO and A2C control policies to determine the optimal drum position across various reactor states, ensuring critical core conditions and symmetric power distribution within all six quadrants. The results demonstrate the exceptional performance of PPO in identifying optimal drum positions, achieving an HPTR value of approximately 1.002 (within the limit of $<$ 1.02) and maintaining criticality within a range of 10 pcm. A2C policies provided lower performance overall compared to PPO where, for certain burnup steps, HPTR constraint was not satisfied. The computational cost for training the RL policy was approximately 60 minutes, while policy control predictions during deployment took around 0.025 seconds for both algorithms. Future extensions of this work include expanding the scope to include additional control variables within the state space (e.g., temperature control), implementing advanced techniques for engineering reward functions to expedite convergence, and transitioning to real-time control for load-following scenarios using transient simulations. 

\section*{Acknowledgment}
The work in this paper was supported by various sources based on the author's contributions: 
\begin{itemize}
    \item Majdi I. Radaideh: Department of Energy, Office of Nuclear Energy Distinguished Early Career Program (Award: DE-NE0009424)
    \item Leo Tunkle: Idaho National Laboratory’s Laboratory Directed Research and Development (LDRD) Program (Award: 24A1081-116FP) under Department of Energy Idaho Operations Office contract no. DE-AC07-05ID14517.
    \item Dean Price: Department of Energy, Office of Nuclear Energy, Integrated University Program Graduate Fellowship.
    \item Kamal Abdulraheem: Michigan Institute of Data Science, Eric and Wendy Schmidt AI in Science Postdoctoral Fellowship.
    \item Linyu Lin: Idaho National Laboratory’s Laboratory Directed Research and Development (LDRD) Program (Award: 24A1081-116FP) under Department of Energy Idaho Operations Office contract no. DE-AC07-05ID14517.
\end{itemize}

This research made use of Idaho National Laboratory's High Performance Computing systems located at the Collaborative Computing Center and supported by the Office of Nuclear Energy of the U.S. Department of Energy and the Nuclear Science User Facilities under Contract No. DE-AC07-05ID14517. 

\section*{CRediT Author Statement}

\begin{itemize}
     \item \textbf{Majdi I. Radaideh}: Conceptualization, Methodology, Software, Validation, Formal analysis, Visualization, Investigation, Funding acquisition, Supervision, Project administration, Writing - Original Draft. 
     \item \textbf{Leo Tunkle}: Methodology, Software, Validation, Formal analysis, Visualization, Investigation, Writing - Original Draft. 
     \item \textbf{Dean Price}: Methodology, Software, Validation, Data curation, Writing - Original Draft. 
     \item \textbf{Kamal Abdulraheem}: Software, Validation, Writing - Review and Edit. 
     \item \textbf{Linyu Lin}:  Methodology, Funding acquisition, Project administration, Writing - Review and Edit.
     \item \textbf{Moutaz Elias}: Conceptualization, Supervision, Writing - Review and Edit. 
 \end{itemize}


\bibliographystyle{elsarticle-num}
\setlength{\bibsep}{0pt plus 0.3ex}
{\small
\bibliography{references}}

\begin{thebibliography}{10}
\expandafter\ifx\csname url\endcsname\relax
  \def\url#1{\texttt{#1}}\fi
\expandafter\ifx\csname urlprefix\endcsname\relax\def\urlprefix{URL }\fi
\expandafter\ifx\csname href\endcsname\relax
  \def\href#1#2{#2} \def\path#1{#1}\fi

\bibitem{shirvan2023uo2}
K.~Shirvan, J.~Buongiorno, R.~MacDonald, B.~Dunkin, S.~Cetiner, E.~Saito, T.~Conboy, C.~Forsberg, Uo2-fueled microreactors: Near-term solutions to emerging markets, Nuclear Engineering and Design 412 (2023) 112470.

\bibitem{papoudakis2019survey}
G.~Papoudakis, F.~Christianos, A.~Rahman, S.~V. Albrecht, \href{http://arxiv.org/abs/1906.04737}{Dealing with non-stationarity in multi-agent deep reinforcement learning}, CoRR abs/1906.04737.
\newblock \href {http://arxiv.org/abs/1906.04737} {\path{arXiv:1906.04737}}.
\newline\urlprefix\url{http://arxiv.org/abs/1906.04737}

\bibitem{mnih2015ogRL}
M.~Volodymyr, K.~Kavukcuoglu, D.~Silver, A.~A. Rusu, J.~Veness, M.~G. Bellemare, e.~a. Alex~Graves, \href{https://doi.org/10.1038/nature14236}{Human-level control through deep reinforcement learning}, Nature 518 (2015) 529–533.
\newline\urlprefix\url{https://doi.org/10.1038/nature14236}

\bibitem{bae2023multiPWR}
J.~Bae, J.~M. Kim, S.~J. Lee, \href{https://doi.org/10.1016/j.net.2023.06.009}{Deep reinforcement learning for a multi-objective operation in a nuclear power plant}, Nuclear Engineering and Technology 55 (2023) 3277–90.
\newline\urlprefix\url{https://doi.org/10.1016/j.net.2023.06.009}

\bibitem{degrave2022magnetic}
J.~Degrave, F.~Felici, J.~Buchli, M.~Neunert, B.~Tracey, F.~Carpanese, T.~Ewalds, R.~Hafner, A.~Abdolmaleki, D.~de~Las~Casas, et~al., Magnetic control of tokamak plasmas through deep reinforcement learning, Nature 602~(7897) (2022) 414--419.

\bibitem{arroyo2022reinforced}
J.~Arroyo, C.~Manna, F.~Spiessens, L.~Helsen, Reinforced model predictive control (rl-mpc) for building energy management, Applied Energy 309 (2022) 118346.

\bibitem{radaideh2021physics}
M.~I. Radaideh, I.~Wolverton, J.~Joseph, J.~J. Tusar, U.~Otgonbaatar, N.~Roy, B.~Forget, K.~Shirvan, Physics-informed reinforcement learning optimization of nuclear assembly design, Nuclear Engineering and Design 372 (2021) 110966.

\bibitem{radaideh2021large}
M.~I. Radaideh, B.~Forget, K.~Shirvan, Large-scale design optimisation of boiling water reactor bundles with neuroevolution, Annals of Nuclear Energy 160 (2021) 108355.

\bibitem{radaideh2023neorl}
M.~I. Radaideh, K.~Du, P.~Seurin, D.~Seyler, X.~Gu, H.~Wang, K.~Shirvan, Neorl: Neuroevolution optimization with reinforcement learning—applications to carbon-free energy systems, Nuclear Engineering and Design 412 (2023) 112423.

\bibitem{chen2022deep}
X.~Chen, A.~Ray, Deep reinforcement learning control of a boiling water reactor, IEEE Transactions on Nuclear Science 69~(8) (2022) 1820--1832.

\bibitem{Kwon1997CNS}
K.-C. Kwon, P.~Jae-Chang, J.~Chul-Hwan, L.~Jang-Soo, J.-Y. Kim, Compact nuclear simulator and its upgrade plan (Nov 1997).

\bibitem{park2022heatup}
J.~Park, T.~Kim, S.~Seong, S.~Koo, \href{https://doi.org/10.1016/j.pnucene.2021.104107}{Control automation in the heat-up mode of a nuclear power plant using reinforcement learning}, Progress in Nuclear Energy 145.
\newline\urlprefix\url{https://doi.org/10.1016/j.pnucene.2021.104107}

\bibitem{leppanen2015serpent}
J.~Lepp{\"a}nen, M.~Pusa, T.~Viitanen, V.~Valtavirta, T.~Kaltiaisenaho, The serpent monte carlo code: Status, development and applications in 2013, Annals of Nuclear Energy 82 (2015) 142--150.

\bibitem{tesauro1995backgammon}
G.~Tesauro, Td-gammon: A self-teaching backgammon program, Applications of neural networks (1995) 267--285.

\bibitem{mnih2013dqn}
V.~Mnih, K.~Kavukcuoglu, D.~Silver, A.~Graves, I.~Antonoglou, D.~Wierstra, M.~Riedmiller, Playing atari with deep reinforcement learning (2013).
\newblock \href {http://arxiv.org/abs/1312.5602} {\path{arXiv:1312.5602}}.

\bibitem{williams1992reinforce}
R.~Williams, \href{https://doi.org/10.1007/BF00992696}{Simple statistical gradient-following algorithms for connectionist reinforcement learning}, Machine Learning 8 (1992) 229--256.
\newline\urlprefix\url{https://doi.org/10.1007/BF00992696}

\bibitem{mnih2016a3c}
V.~Mnih, A.~P. Badia, M.~Mirza, A.~Graves, T.~Lillicrap, T.~Harley, D.~Silver, K.~Kavukcuoglu, \href{https://proceedings.mlr.press/v48/mniha16.html}{Asynchronous methods for deep reinforcement learning}, in: M.~F. Balcan, K.~Q. Weinberger (Eds.), Proceedings of The 33rd International Conference on Machine Learning, Vol.~48 of Proceedings of Machine Learning Research, PMLR, New York, New York, USA, 2016, pp. 1928--1937.
\newline\urlprefix\url{https://proceedings.mlr.press/v48/mniha16.html}

\bibitem{sb3}
A.~Raffin, A.~Hill, A.~Gleave, A.~Kanervisto, M.~Ernestus, N.~Dormann, \href{http://jmlr.org/papers/v22/20-1364.html}{Stable-baselines3: Reliable reinforcement learning implementations}, Journal of Machine Learning Research 22~(268) (2021) 1--8.
\newline\urlprefix\url{http://jmlr.org/papers/v22/20-1364.html}

\bibitem{schulman2015trpo}
J.~Schulman, S.~Levine, P.~Abbeel, M.~Jordan, P.~Moritz, \href{https://proceedings.mlr.press/v37/schulman15.html}{Trust region policy optimization}, in: F.~Bach, D.~Blei (Eds.), Proceedings of the 32nd International Conference on Machine Learning, Vol.~37 of Proceedings of Machine Learning Research, PMLR, Lille, France, 2015, pp. 1889--1897.
\newline\urlprefix\url{https://proceedings.mlr.press/v37/schulman15.html}

\bibitem{schulman2017ppo}
J.~Schulman, F.~Wolski, P.~Dhariwal, A.~Radford, O.~Klimov, Proximal policy optimization algorithms (2017).
\newblock \href {http://arxiv.org/abs/1707.06347} {\path{arXiv:1707.06347}}.

\bibitem{schulman2018gae}
J.~Schulman, P.~Moritz, S.~Levine, M.~Jordan, P.~Abbeel, High-dimensional continuous control using generalized advantage estimation (2018).
\newblock \href {http://arxiv.org/abs/1506.02438} {\path{arXiv:1506.02438}}.

\bibitem{price2023thermal}
D.~Price, N.~Roskoff, M.~I. Radaideh, B.~Kochunas, Thermal modeling of an evinci™-like heat pipe microreactor using openfoam, Nuclear Engineering and Design 415 (2023) 112709.

\bibitem{openaigym}
G.~Brockman, V.~Cheung, L.~Pettersson, J.~Schneider, J.~Schulman, J.~Tang, W.~Zaremba, Openai gym (2016).
\newblock \href {http://arxiv.org/abs/arXiv:1606.01540} {\path{arXiv:arXiv:1606.01540}}.

\bibitem{radaideh2024demonstration}
M.~I. Radaideh, D.~Price, K.~Abdulraheem, M.~Elias, Demonstration of microreactor reactivity control with reinforcement learning, in: International Conference on Physics of Reactors (PHYSOR 2024), American Nuclear Society, 2024, pp. 1427--1436.

\bibitem{chollet2015keras}
F.~Chollet, et~al., Keras, \url{https://keras.io} (2015).

\bibitem{radaideh2020surrogate}
M.~I. Radaideh, T.~Kozlowski, Surrogate modeling of advanced computer simulations using deep gaussian processes, Reliability Engineering \& System Safety 195 (2020) 106731.

\bibitem{radaideh2020neural}
M.~I. Radaideh, C.~Pigg, T.~Kozlowski, Y.~Deng, A.~Qu, Neural-based time series forecasting of loss of coolant accidents in nuclear power plants, Expert Systems with Applications 160 (2020) 113699.

\end{thebibliography}

\end{document}